\documentstyle[aps,a4,twocolumn,psfig]{revtex}

\begin{document}
\title{Emergence of Cooperation and Organization in an Evolutionary
Game}
\author{D. Challet and Y.-C. Zhang\\{\small Institut de Physique 
Th\'eorique, Universit\'e de Fribourg, 1700 Fribourg, Switzerland}}
\maketitle

\abstract{\it A binary game is introduced and analysed. N players have to choose
one of the two sides independently and those on the minority side win. Players
uses a finite set of ad hoc strategies to make their decision, based on the past
record. The analysing power is limited and can adapt when necessary. Interesting
cooperation and competition pattern of the society seem to arise and to be 
responsive to the payoff function.}
\vspace{3ex}

Most current economics theories are deductive in origin. One assumes
that each participant knows what is best for him given that all other
participants are equally intelligent in choosing their best
actions. However, it is recently realised that in the real world the
actual players do not have the perfect foresight and hindsight, most
often their actions are based on trial--and--error inductive thinking,
rather than the deductive rationale assuming that there are underlying first
principles. Whether deductive or inductive thinking is more relevant
is still under debate [1]. 

Evolutionary games have also been studied within the standard framework
of game theory [2]. However it is recently pointed out that the approach
traditionally used in economics is not convenient to generalise to include
irrationality, and an alternative langevin type equation is proposed [3].
As physicists we would like to view a game with a large number of players 
a statistical system, we need to explore new approaches in which emerging 
collective phenomena can be better appreciated.
One recent approach using bounded rationality is
particularly inspiring, put forward by B. Arthur in his {\it El Farol} 
bar problem [4]. Following the similar philosophy,
in this work we propose and study a simple evolutionary game.

Let us consider a population of $N$ (odd) players, each has some
finite number of strategies $S$. At each time step, everybody has to
choose to be in side $A$ or side $B$. The payoff of the game is to
declare that after everybody has chosen side independently, those who
are in the minority side win. In the simpliest version, all winners
collect a point. The players make decisions based on the common
knowledge of the past record. We further limit the record to contain
only yes and no e.g. the side $A$ is the winning side or not, without
the actual attendance number. Thus the system's signal can be represented by a
binary sequence, meaning $A$ is the winning side (1) or not (0).

Let us assume that our players are quite limited in their analysing
power, they can only retain last $M$ bits of the system's signal and
make their next decision basing only on these $M$ bits. Each player
has a finite set of strategies. A strategy is defined to be the next action (to
be in $A$ or $B$) given a specific signal's M bits. The example of one
strategy is illustrated in table 1 for M=3.

\begin{center}
\begin{tabular}{|c|c|}\hline
        signal&prediction\\ \hline
        000&1\\ \hline
        001&0\\ \hline
        010&0\\ \hline
        011&1\\ \hline
        100&1\\ \hline
        101&0\\ \hline
        110&1\\ \hline
        111&0\\ \hline
\end{tabular}
\end{center}

There are 8 ($=2^M$) bits we can assign to the right side, each configuration
corresponds a distinct strategy, this makes the total number of
strategy to be $2^{2^M}=256$. This is indeed a fast increasing number,
for $M=2$, $3$, $4$, $5$ it is $16$, $256$, $65536$, $65536^2$. We
randomly draw $S$ strategies for each player, and some strategies
maybe by chance shared. However for moderately large $M$, the
chance of repetition of a single strategy is exceedingly
small. Another special case is to have all 1's (or 0's) on the RHS of the
table, corresponding to the fixed strategy of staying at one side no matter
what happens.

Let us analyse the structure of this Minority's game to see what to
expect. Consider the extreme case where only one player takes a side,
all the others take the other. The lucky player gets a reward
point, nothing for the others. Equally extreme example is that when
$(N-1)/2$ players in one side, $(N+1)/2$ on the
other. From the society point of view, the second situation is
preferable since the whole population gets $(N-1)/2$ points
whereas in the first example only one point---a huge waste. Perfect
coordination and timing would approach the 2nd, disaster would be the
first example. In general we expect the population to behave between
the above two extremes.

This binary game can be easily simulated for a large population of
players. Initially, each player draws randomly one out of his $S$
strategies and use it to predict next step, an artificial signal of
$M$ bits is also given. All the $S$ strategies in a player's bag can
collect points depending if they would win or not given the $M$ past
bits, and the actual outcome of the next play. However, these points
are only $virtual$ points as they record the merit of a strategy as if it
were used each time. The player uses the strategy having the highest
accumulated points (capital) for his action, he gets a real point only
if the strategy used happens to win in the next play.
\begin{center}
\begin{figure}
\psfig{figure=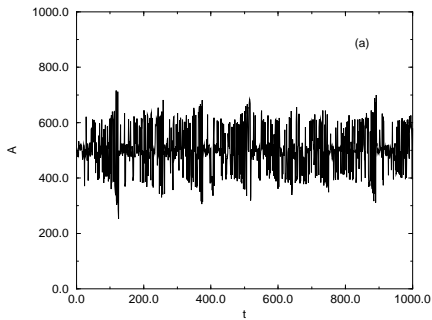,width=6cm}
\psfig{figure=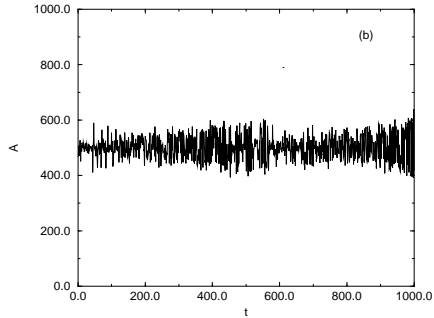,width=6cm}
\psfig{figure=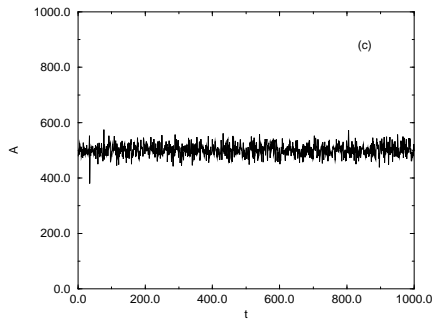,width=6cm}
\caption{Actual number of attendance at the side A against time, for a population of 1001 players,
        having brain size of 6, 8 and 10 bits (a),(b),(c)}
\end{figure}
\end{center}\vspace{-3ex}
In Figs. 1 we plot the actual number of attendance at the side A, for
a population of 1001 players, having various brain sizes (i.e. $M$
bits). As one may expect, that the temporal signal indeed fluctuates
around the 50 $\%$. Whoever takes the side $A$ wins a point at a given
time step when the signal is below 501. The precise number is
not known to the players, they only know if a side is  winning or not,
after their bet is made. Note that large fluctuations imply large waste
since still more players could have taken the winning side without harm
done to the others. On the other hand, smaller fluctuations imply more
efficient usage of available resources, in general this would require
coordination and cooperation --- which are not built-in
explicitly. We see
that the population having larger brains (i.e. M larger) cope with
each other better : the fluctuation are indeed in decreasing order for
ever increasingly ''intelligent'' players (i.e. $M=6$, $8$,
$10$). Remarkable is that each player is by definition selfish, not
considerate to fellow players, yet somehow they manage to better
somewhat share the limited available resources.

Let us remark that the very simplest strategy by playing randomly is
not included here, for generating random numbers more bits are needed. 
In a perfect timing, the average gain in the population would be $1/2$
per play. Waste is proportional to fluctuation's amplitude hence the
average gain is always below $1/2$ in reality. Since the game is
symmetrical in $A$ and $B$, one may be tempted to use the simple
strategy to stay at $A$ or $B$, hoping to get exactly $1/2$ gain. Let
us mention if this strategy indeed rewards $1/2$ gain on average, many
would imitate. Let us say there is a group sitting at $A$ no matter
what signal is shown (this is included in the strategy space). The active
players  will soon recognise that they win less often choosing $A$ than
$B$. In fact, for them the game is no longer symmetrical and they will
adopt accordingly so that the apparent advantage disappears for those
sitting at one side fixed. This is similar to the arbitrage
opportunities in finance: any obvious advantage will be arbitraged
away --- no easy ``risk-free way'' to make a living both for our
players and those in the real world.

\begin{center}
\begin{figure}
\psfig{figure=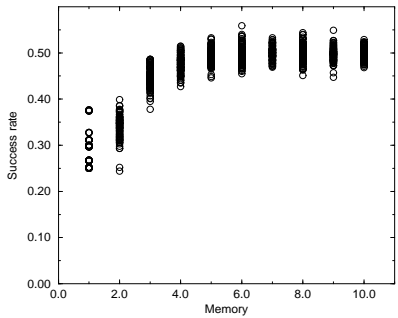,width=6cm}
\caption{ Success rate of a mixed population players against their memory ($N=1001$, $S=5$)}
\end{figure}
\end{center}

The advantage of the larger brain sizes over the smaller ones can be
better appreciated inspecting Fig. 2. Identical parameters ($N=1001$,
$S=5$) for a mixed population having $M=1,\cdots,10$. 
We thus force unequally equiped players to
play together. One may fear that the ''poorly'' brained players may
get exploited by the more powerfully brained ones. Indeed this is the
case.
We plot the average
gain per time step after a long time. We see that within a sub-population
(same $M$) there are better and worse performers. We have noticed that better
players do not necessarily stay that way for a long time, but
exceptions exist. For $M=1$, there appears fewer points, since there are
more degeneracies. As a group the more intelligent players gain more
and the spread between the rich and the poor is smaller, even though
the in-fighting among them is more intensified. Remark that above a
certain size ($M\approx 6$) the average performance of a population
appears to saturate, further increasing the brain size does not
seem to improve more. This is due to the simple structure of
this version of the game, there is nothing more to gain. Recall that
only most crude information is transmitted to the players, i.e. only
yes and no, not the exact attendance number. More precise information
would necessitate more analysing power, more complicated payoff functions
and games also provides incentives to develop more sophisticated
brains. However in the present work, we stick to the binary functions
and will report more complicated applications using neural
networks elsewhere.

\begin{center}
\begin{figure}
\psfig{figure=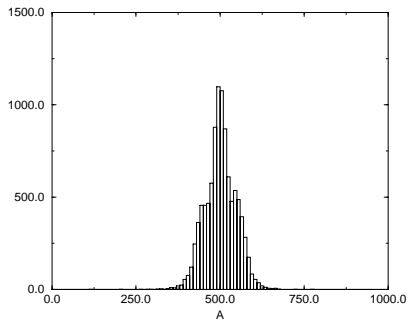,width=6cm}
\caption{Histograph of the attendance of $A$ ($N=1001$, $M=8$, $S=5$)}
\end{figure}
\end{center}

Of course the game is symmetrical for $A$ and $B$. This can be
observed in Fig. 3, where the histograph shows
the attendance of $A$ (hence $B$ is the mirror image at the point
$N=501$). B. Arthur's {\it El Farol} problem uses $60\%$ rule and does not
rise new questions and results appear to be similar.

One may argue that our payoff function is too simple, i.e. a step
function without differentiating a ``good'' minority from a ``bad''
one. Let us consider the payoff function $N/x-2$, i.e. these
many (nearest integer values) points awarded to every player choosing the
minority side, the number of  winning players  being $x<N/2$. 
Clearly this structure favours
smaller minority. This is like in lottery you would like to be on the
winning side, but even better you are alone there. The
players thus face an extra type  of competition, a winner would prefer
less fellow winners in company. If for instance a
player wins on a mediocre play, his winning strategies are hardly
enhanced with respect to not winning at all. Globally the population
($N=1001$, $M=4$) respond to having a histograph Fig. 4 with two
peaks. Although the jackpot (winning alone) is very appealing, this
is very unlikely to happen since the fellow players are just as
intelligent. The players need a sizeable gain to get motivation
to win. The appears to be a compromise that they effectively (not through
any enforceable agreement) agree to show up on the
minority side a smaller number of players. What is remarkable here is
that entropy, i.e. the most likely configuration, does not favour the
Fig 4. distribution. The players manage to defy entropy, in other
words to get themselves organised to occupy less unlikely
configurations.

\begin{center}
\begin{figure}
\psfig{figure=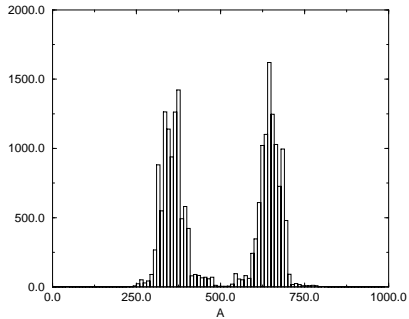,width=6cm}
\caption{Histograph of the attendance of $A$ for a $N/x-2$ payoff ($N=1001$, $M=4$, $S=5$)}
\end{figure}
\end{center}

One may inquire what happens if the players are provided with a bigger
``idea bag'' with more alternative strategies. In Fig. 5 we show the
results for various populations (N=$1001$, $M=5$) with
$S=2,\:3,\cdots,9$. We see that in general with increasing number of
alternatives the players tend to perform worse. What happens is that
the players tend to switch strategies oftener and more likely to get
``confused'', i.e. some outperforming strategy may distract the player's
attention, after being chosen turns out to be underperforming. We
recognise that this has also to do with the observation time, currently a
player switches immediately if another strategy has one virtual point
more than that in use. If a higher threshold is set, then the
hinderance by increasing number of alternatives can be in part avoided. 
In the
neural network version of our game, just one network (with adjustable
weights) is given to a player. Let us recall that in a recent study,
Borkar {\em et al} [5] have proven that in an evolutionary game players tend to 
specialise in a single
strategy, even though alternatives exist.

\begin{center}
\begin{figure}
\psfig{figure=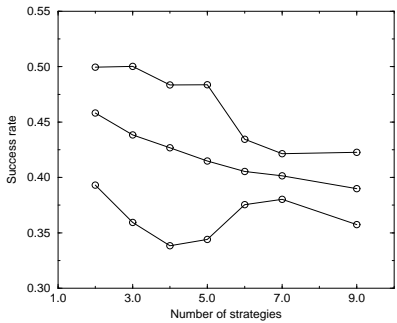,width=6cm}
\caption{Success rate of the best, worst and average players against the 
number of strategies ($N=1001$, $M=5$)}
\end{figure}
\end{center}

\begin{center}
\begin{figure}
\psfig{figure=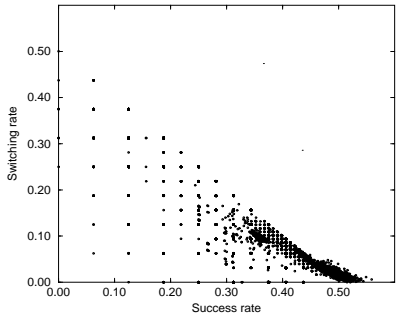,width=6cm}
\caption{Switching rate against the succes rate for various populations}
\end{figure}
\end{center}

In Fig. 6 we plot the switching rate against the succes rate for
various populations. The general tendance that the oftener one
switches, less successful one would end up. The phase space seems
to be highly fragmented and many substructures appear, attributable to
the binary nature of our game.

\begin{center}
\begin{figure}
\psfig{figure=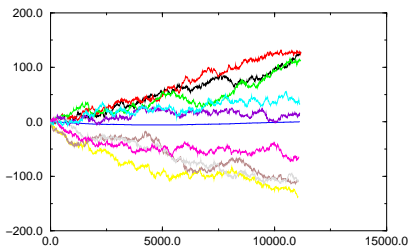,width=6cm}
\caption{Performance record of the 3 best, the 3 worse and 3 randomly chosen players ($N=1001$, $M=10$, $S=5$)}
\end{figure}
\end{center}

It is also instructive to follow the performance record. In
Fig 7., we select 3 top players, 3 bottom players and 3 randomly
chosen players. They are chosen at the last time step and we trace
back their past record. Their capital gains are scaled such that the
average gain (over the population) appears in an almost horizontal line. 
We see that the general
tendance for the best and worst players are rather consistent even though
setbacks for the best and bursts for the worst do occur. Notice that
the gap between the rich and the poor appears to increase linearly with
time, though reversion is possible but the poor players in general are
doomed to stay poor.

\begin{center}
\begin{figure}
\psfig{figure=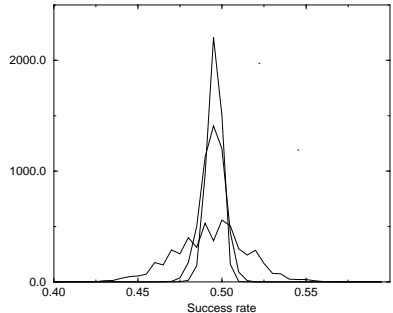,width=6cm}
\caption{Different distributions of the average value of all the strategies with increasing iterations numbers (1000, 5000
and 10000), showing that all strategies are equivalent in the $t\rightarrow \infty$ limit }
\end{figure}
\end{center}

Another result enhances this conclusion : one may blame bad players for
their bad strategies. In order to check whether there are really good
and bad strategies, we plot the virtual gains of all the strategies in
the population. In Fig 8 we see three different distributions of the
average (time) gains. The longer the time the more concentrate is the
distribution, indicating that relative values of the strategies are
about the same. 
Indeed it can be analyticaly shown that all the strategies are equivalent
to each other, since our game is symmetrical in $A$ and $B$.
So the bad
player are bad because they have used the strategies inopportunely and are
unlucky, also their specific composition is to blame. Note that a player
is only distinguished from others by this composition, if two players
having the same composition, they are clone sisters. In that case
initial conditions can still set them apart and they may know
different fortunes only in the beginning.

The above discussion calls for a genetic approach in which the poor
players are regulary weeded out from the game and new players are 
introduced to replace the eliminated ones. This genetic approach is already applied to
a prototype model [6] for stock or currency trading, mimicking what happens in the real
market. Let us consider our
minority game generalized to include the Darwinist selection : the
worst player is replaced by a new one after some time steps,
the new player is a clone of the best player, i.e. it inherits all
the strategies but with corresponding virtual capitals reset to
zero. This is analogous to a new born baby, though having all the 
predispositions
from the parents, does not inherit their knowledge.

\begin{center}
\begin{figure}
\psfig{figure=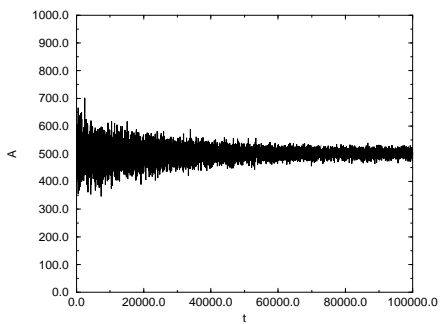,width=6cm}
\caption{Temporal attendance of $A$ for the genetic approach showing a learning process}
\end{figure}
\end{center}

To keep a certain diversity we introduce a mutation possibility in
cloning. We allow one of the
strategies of the best player to be replaced by a new one. Since strategies
are not just recycled among the players any more, the whole strategy
phase space is available for selection. We expect this population is
capable of ``learning'' since self-destructive, obviously bad players
are weeded out with time, fighting is among so-to-speak the best
players. Indeed in Fig 9 we observe that the learning has emerged in
time. Fluctuations are reduced and saturated, this implies the average
gain for everybody is improved but never reaches the ideal limit.
What would happen if no mutation is allowed and cloning is perfect?
Eventually population is full of the clone copies of the best player,
each may still differ in their decision since the virtual capitals
in their idea-bag can be different. In Fig. 10 we plot the performance
of such a ``pure" population, there appears tremendous waste and all
strange things go loose. Indeed, the results from inbreeding look 
rather incestous.

\begin{center}
\begin{figure}
\psfig{figure=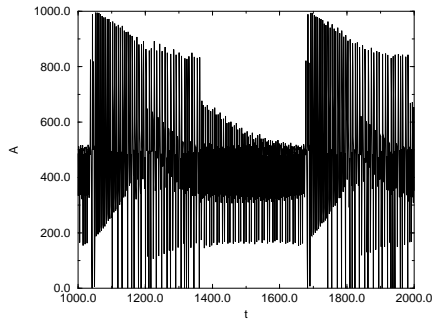,width=6cm}
\caption{Temporal attendance of $A$ of an ``pure'' population}
\end{figure}
\end{center}

As the last experiment we start the population very ``simple-minded'',
say $M=2$. We allow in the cloning process mentionned above an
additional feature that a bit of memory can be added or substracted
for the cloned new player, with a small probability. We want
to be sure that the rules are such that this structural mutation is
strictly neutral, i.e  does not favour bigger brains over the smaller
ones, we leave that to the invisible hand of evolution to
decide. Indeed something remarkable takes place: in Fig. 11 we plot
the average brain size in the population started with $M=2$, for a
population of $N=101$ and $N=1001$. The
temporal record shows that there is an ``arm race'' among the
players. We know by now that the more brain power leads to advantage,
so in the evolution of survival-of-the-fittest the players develop
bigger brains to cope with ever aggressive fellow-players. However
such an evolution appear to saturate and the ``arm race'' to settle at
a given level. The saturation values are not universal, having to do with
the time intervals of reproduction. In general the larger brains need longer time 
to learn. Larger population ($N=1001$) needs more powerful
brains to sustain the apparent equilibrium than the smaller population
($N=1001$), also the learning rate (the slope in Fig.11) is smaller. 
We mention {\em en passant} that population's brain sizes 
do not concentrate on one value, only average value
is plotted. Some players manage to
make do quite happily with a relatively small brains. 

\begin{center}
\begin{figure}
\psfig{figure=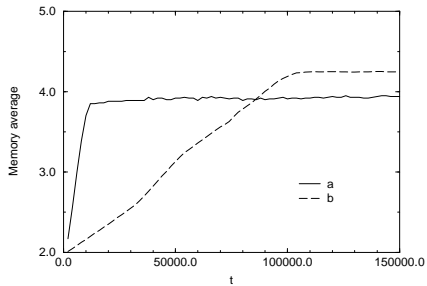,width=6cm}
\caption{Temporal record of the memory average of a starting from $M=2$ population for $N=101$ (a) and $N=1001$
(b) ($S=5$)}
\end{figure}
\end{center}
\vspace{-7ex}
{\bf Conclusions}. What can we learn from these simple numerical experiments? First of 
all the economical behavior in the real world seems to call for a general
approach to systematically study the evolutionary nature of games. There are a few
most relevant questions to address: 1) given each agent's selfishness what is
his cooperative and cognitive skills in the course of competition ? 2) What is
the emerging collective behavior that is the society's performance without an
enforceable authority ? 3) How can our {\it visible} hand modify the rules of the game
(payoff functions) such that global response may appear more cooperative ? 4) 
How does evolution puts its {\it invisible} hand to work?
Clearly our study is just a beginning to answer all these. What we have presented in this
work is not just an oversimplied model, but a general approach to ask the right
questions. This approach, as the reader can readily convince himself, is very
open to all sorts of variation. It is easy to include other
situation-motivated payoff functions and game structures, there are
qualitatively new questions to be asked when more realistic games are studied.
It is a theoretical physicist's dream to have an Ising type model, though
oversimplified, and yet to capture some essential points of the real
world. Our minority game may be indeed the simplest of the kind. 

Our model is by design without fundamentals and insider information. Players
are forced to fight each other. With the Darwinism included, everyone has to
keep improving in order to survive---the {\it red queen} effect. 
Unlike some examples in standard game
theory, there is no commonly accepted optimal strategy (analogous to physical
systems without obvious ground states). A rational approach is helpless here.
Yet the emerging society appears to have a certain organisation. Even though
the players care only their own gain, cooperation and timing does seem to
spontaneously arise. Note that our learning mechanism is different from the traditional
neural network studies, where a {\it pre-assigned} task like a pattern is given and
performance is measured on how precise the original is restored. Here the task is
self-appointed and no ending is defined.

We may even speak of the emergence of intelligence. If the analysing power of
the players can be adapted to the increasingly challenging task (survival
amongst ever aggressive fellow players and larger number of players), the
populations seem to evolve more equipped, larger brains appear to dominate and
available resources are better exploited, i.e. less fluctuation and waste in
attendance number. This is not unsimilar to the study of the prebiotic
evolution: in the promodial soup only very simple organisms exist. Evolution
allows these organisms to add one new feature (or reduce an existing one) from
time to time. More complex organisms tend to cope with the survival task better and
more and more refined organisms spontaneously appear out of the
monotonous soup [7]. 

We thank Matteo Marsili for helpful conversations. This work has been supported in part by the
Swiss National Foundation through the Grant No. 20-46918.96.

{\bf References :} 

[1] {\em The Economy as an Evolving Complex System}, 
ed. P.W. Anderson, K. Arrow and D. Pines, Redwood City, Addison-Wesley
Co.,1988 

For a recent example

[2] J. W. Weibull, {\em Evolutionary Game Theory}, MIT Pres, Cambridge, 1995

[3] M. Marsili and Y.-C. Zhang, {\em Fluctuations around Nash Equilibria in Game Theory}, Physica A, to appear

[4] W. B. Arthur, {\em Inductive Reasoning and Bounded Rationality}, Am. Econ. Assoc. Papers and Proc. 84, pp. 406-411, 1994

[5] V. S. Borkar, S. Jain, G. Rangarajan, {\em Dynamics of Individual Specialization and Global Diversification in
Communities}, preprint, {\em Bangalore}, IISc-CTS-3/97

[6] G. Caldarelli, M. Marsili and Y.-C. Zhang, {\em A prototype Model of stock
exchange}, Europhysics Letters, to appear

[7] Y.-C. Zhang, unpublished; {\em Quasispecies Evolution of Finite Population} Phys. Rev. E 55, pp. R3815, 1997

\end{document}